\documentclass[10pt]{article}
\usepackage[OE]{express}

\begin{document}
\title{Anti-reflection coating design for metallic terahertz meta-materials}

\author{Matteo Pancaldi\authormark{1,2}, Ryan Freeman\authormark{3}, Matthias Hudl\authormark{1}, Matthias C. Hoffmann\authormark{4}, Sergei Urazhdin\authormark{3}, Paolo Vavassori\authormark{2,5} and Stefano Bonetti\authormark{1,*}}

\address{\authormark{1}Department of Physics, Stockholm University, 106 91 Stockholm, Sweden\\
\authormark{2}CIC nanoGUNE, E-20018 Donostia-San Sebastian, Spain\\
\authormark{3}Emory University, Atlanta, GA 30322, USA\\
\authormark{4}SLAC National Accelerator Laboratory, Menlo Park, CA 94025, USA\\
\authormark{5}IKERBASQUE, Basque Foundation for Science, E-48013 Bilbao, Spain}

\email{\authormark{*}stefano.bonetti@fysik.su.se}

\begin{abstract}
We demonstrate a silicon-based, single-layer anti-reflection coating that suppresses the reflectivity of metals at near-infrared frequencies, enabling optical probing of nano-scale structures embedded in highly reflective surroundings. Our design does not affect the interaction of terahertz radiation with metallic structures that can be used to achieve terahertz near-field enhancement. We have verified the functionality of the design by calculating and measuring the reflectivity of both infrared and terahertz radiation from a silicon/gold double layer as a function of the silicon thickness. We have also fabricated the unit cell of a terahertz meta-material, a dipole antenna comprising two 20-nm thick extended gold plates separated by a 2 $\mu$m gap, where the terahertz field is locally enhanced. We used the time-domain finite element method to demonstrate that such near-field enhancement is preserved in the presence of the anti-reflection coating. Finally, we performed magneto-optical Kerr effect measurements on a single 3-nm thick, 1-$\mu$m wide magnetic wire placed in the gap of such a dipole antenna. The wire only occupies 2\% of the area probed by the laser beam, but its magneto-optical response can be clearly detected. Our design paves the way for ultrafast time-resolved studies, using table-top femtosecond near-infrared lasers, of dynamics in nano-structures driven by strong terahertz radiation.
\end{abstract}


\bibliographystyle{unsrt}

\section{Introduction}
Following the development of intense, coherent laser-based sources of terahertz radiation \cite{hoffmann11}, the past decade has witnessed an increased interest in the use of this type of radiation to coherently control the properties of materials on the sub-picosecond time scale. Terahertz photons, with energies in the meV range, can drive nonlinear dynamics without significantly increasing the entropy of the system \cite{trigo08,liu12,daranciang12,mankowsky14,staub14,beaud14,kubacka14,dakovski15,bonetti16,
dean16,henighan16,grubel16,chen16}. In the field of condensed matter physics, the investigations of ultrafast dynamics driven by strong terahertz fields are frequently performed using terahertz-pump (usually in the 1 - 10 THz range, 300 to 30 $\mu$m in wavelength) and visible or near-infrared probing light (typically a sub-100 fs pulse).

To study the effects in the strong-field limit, the strength of the terahertz field can be locally enhanced exploiting near-field effects in meta-materials \cite{chen06,razzari11,werley12,liu12,zhang15,savoini16,kozina17}, which typically consist of micrometer-sized metallic structures deposited on the sample surface. However, since the area of the sample is often significantly smaller than the area of metallic structures in meta-materials, the reflectivity in the visible or near-infrared frequency range of the probe is dominated by the latter. As a consequence, it is extremely challenging to isolate the sample response, despite the enhancement provided by the meta-material. This problem can be mitigated by using dielectric and absorbing coatings, for instance  to enhance the magneto-optical activity in magnetic thin films, and to reduce the background reflections \cite{balasubramanian88,atkinson91,qureshi04,qureshi05,barman06,wang07}. This solution greatly boosts the signal up to a point where single nano-structures can be measured. The drawback of this approach is that it imposes constraints on the choice of layers underneath the target structure. This limitation can become crucial if these underlayers are utilized  to tune the important properties of the studied thin films. In this case, a more suitable solution is to deposit an anti-reflection (AR) coating only on the metal structures forming the meta-material, to minimize the reflection from those areas, which is the main factor affecting the strength of the measured signal. At the same time, the AR layer should not perturb the terahertz radiation that still needs to be enhanced by the metal layers.

In this work, we propose a simple, but until now unexplored, single-layer anti-reflection coating design that can be implemented on arbitrary meta-material structures comprising highly conducting and reflective metallic layers. The coating suppresses the near-infrared reflection typically utilized to probe the response of the sample, without noticeably affecting the terahertz radiation at much larger wavelengths. We have performed transfer matrix method calculations, as well as measurements of the reflectivity both in the near-infrared and terahertz range, to demonstrate the functionality of our design. We have also investigated, using time-domain finite element simulations, the near-field enhancement properties of a dipole antenna - a template for the terahertz meta-materials - covered with the anti-reflection coating. Finally, we experimentally measured the magneto-optical Kerr effect from a magnetic wire placed in the gap of the antenna.

\section{Anti-reflection coating design}
\begin{figure}[t]
\centering\includegraphics[width=0.8\textwidth]{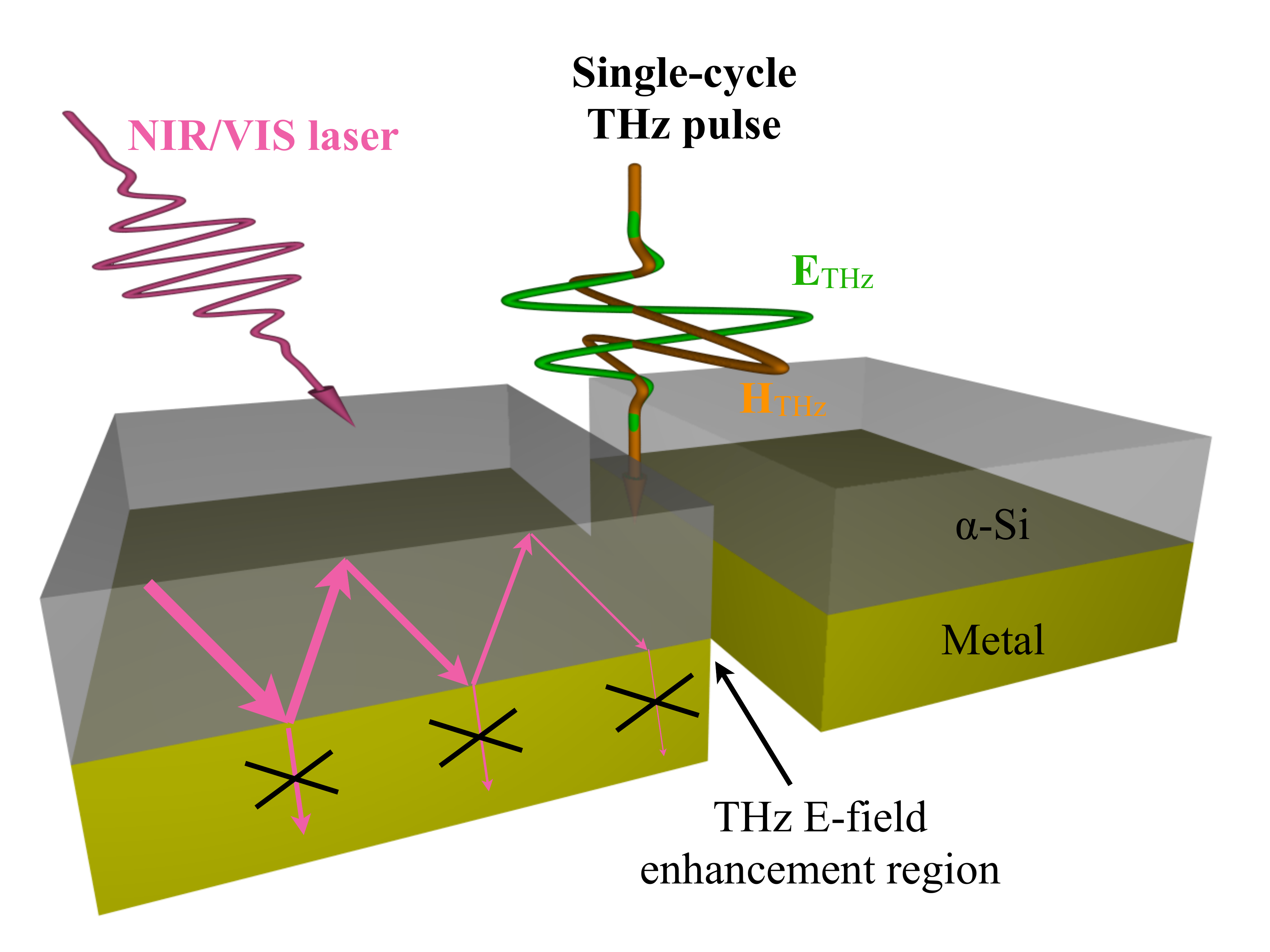}
\caption{Design of the dipole antenna for terahertz near-field enhancement in the gap between two metallic electrodes, covered with an anti-reflection coating for near-infrared and visible radiation. A single-cycle of the terahertz field, with the suitable polarization for the optimal coupling to the antenna, is sketched. The pink arrows schematically show the working principle of the anti-reflection coating for a metal, where destructive interference (zig-zag arrows within the top layer) is combined with the dielectric losses to compensate for the forbidden transmission through the metallic electrodes (crossed-out arrows in the metal layer), as described in detail in the text.}
\label{fig1}
\end{figure}

Terahertz meta-materials can be formed by depositing metallic (typically gold) layers that can locally enhance the electromagnetic field of incident radiation. One of the simplest realization of such structure consists of two metallic strips separated by a small gap, i.e. a dipole antenna \cite{biagioni12}. For a suitable geometry of the antenna and polarization of the incident radiation, opposite charges can be induced by the electromagnetic field at the opposite edges of the gap, producing a strong local enhancement of electric field within the gap. Intuitively, but incorrectly, this charge motion is often attributed to the current driven by the electric field. However, the correct explanation is that the local electric field enhancement in the gap is caused by the screening of the \emph{magnetic} field, which induces a current flow in the metal, in the direction orthogonal to it (parallel to the electric field). This current flow, known as the eddy current, can penetrate within the skin depth of the material (75 nm at 1 THz for gold). In contrast, the \emph{electric} field component of the radiation is screened virtually instantaneously at the surface of the conductor by the charge redistribution, and at terahertz frequencies provides a negligible contribution to the net current flow in the bulk of the material, even in non-ideal metals.

In the standard AR coatings, designed to minimize the reflection from dielectric materials, one exploits the phenomenon of destructive interference of the waves reflected at the two interfaces, to cancel the total electric field that propagates in the backward direction. Since the energy of the electromagnetic wave is conserved, the transmission through the dielectric is maximized. However, this mechanism cannot be implemented for coatings on metals, since the wave cannot propagate through the metal, and hence the reflection cannot be eliminated.

For an AR coating to work for a metal, it is necessary to create destructive interference (to suppress Fresnel reflections), and to simultaneously absorb the radiation, as shown schematically in Fig.~\ref{fig1}. In other words, the dielectric layer needs to be sufficiently lossy in the visible/near-infrared region, so that the wave decays after multiple reflections at the interfaces. This idea was proposed decades ago by Hass et al. \cite{hass56}, who demonstrated that lossy double dielectric layers can suppress the reflectivity of aluminum and copper in the visible range, while maintaining high-reflectivity in the mid-infrared range, up to the wavelength of 10 $\mu$m. Moreover, they also highlighted the fact that absorption in single-layer AR coatings is necessary to reduce the high reflectance of metals in the visible range. In a later related work by Yoshida \cite{yoshida79}, a single-layer AR coating for metals was described mathematically. He first considered a non-absorbing dielectric layer with real refractive index $n_1>1$ and thickness $d_1$, deposited on top of a metallic substrate characterized by $\tilde{n}_2=n_2+ik_2$. The reflectance $R$ for the monochromatic light with wavelength $\lambda$, impinging at normal incidence on the three-layers stack composed of the air ($n_0=1$), the non-absorbing dielectric coating, and the metal substrate, is given by \cite{yoshida79}
\begin{equation}
R~=~\left| \dfrac{r_{01}+r_{12}\exp\left(2i\delta_1\right)}{1+r_{01}r_{12}\exp\left(2i\delta_1\right)} \right|^2,
\end{equation}
where $r_{01}=\left(1-n_1\right)/\left(1+n_1\right)$ is the Fresnel reflection coefficient for the air-dielectric interface, $r_{12}=\left(n_1-\tilde{n}_2\right)/\left(n_1+\tilde{n}_2\right)$ is the Fresnel reflection coefficient for the dielectric-metal interface, and $\delta_1=2\pi n_1d_1/\lambda$. The reflectance reaches a minimum when \cite{park64},
\begin{equation}
n_1d_1~=~\dfrac{\lambda}{2}\left[(m+1)-\dfrac{\alpha_{12}}{2\pi}\right],
\label{optpath}
\end{equation}
where $m$ is an integer value, and
\begin{align}
R_{min}~&=~\left(\dfrac{r_{01}+\rho_{12}}{1+r_{01}\rho_{12}}\right)^2, \\
{\rho_{12}}~&=~\left|r_{12}\right|~=~\left[\dfrac{\left(n_1-n_2\right)^2+{k_2}^2}{\left(n_1+n_2\right)^2+{k_2}^2}\right]^{\frac{1}{2}}, \\
\alpha_{12}~&=~\mathrm{Arg}\left(r_{12}\right)~=~\arctan\left(\dfrac{-2n_1k_2}{{n_1}^2-{n_2}^2-{k_2}^2}\right). \label{alpha_12}
\end{align}
The minimum reflectance $R_{min}$ is zero when $\rho_{12}=|r_{01}|$, giving
\begin{equation}
n_1~=~\left(n_2+\dfrac{{k_2}^2}{n_2-1}\right)^{\frac{1}{2}}.
\label{n1_noabs}
\end{equation}
If $k_2=0$, which corresponds to a dielectric coating on top of a dielectric substrate, Eq.~\eqref{n1_noabs} gives $n_1={n_2}^{1/2}$. Since $n_2>n_1$, the solution for $\alpha_{12}=\arctan\left(0\right)$ in Eq.~\eqref{alpha_12} should be $\alpha_{12}=\pi$, because of the $\pi$ phase shift introduced by the dielectric coating-dielectric substrate interface in this case. Thus, Eq.~\eqref{optpath} gives $n_1d_1=\lambda/4$ for $m=0$, defining a \emph{quarter-wave} coating, in which the reflectance is minimized by the destructive interference in the coating layer. Moreover, Eq.~\eqref{n1_noabs} imposes a constraint on the values of $n_2$ and $k_2$, since $n_1>1$. The effect of this constraint is that, according to Yoshida \cite{yoshida79}, ``zero reflection cannot be achieved with a single dielectric film coating for metals'' with large extinction coefficient $k\gtrapprox3$, such as silver and gold. However, in this case, zero reflection can be obtained by allowing the dielectric coating to be slightly absorbing.

In the following, we experimentally confirm that the reflection from gold, and hence from any good metal, can be suppressed by using a single layer of sputtered amorphous silicon ($\alpha$-Si). In the visible/near-infrared range, a thin $\alpha$-Si film acts as a dielectric with a relatively large imaginary part of the refractive index, since the electronic states are not characterized by well-defined momentum, enhancing the radiation absorption in $\alpha$-Si as compared to its crystalline form \cite{adachi00}. On the other hand, low absorption in the terahertz range  ($\lambda\sim100$ $\mu$m), and the small thickness compared to the radiation wavelength, make these layers practically invisible, thus maintaining the high-reflectivity characteristics of gold in this range.

\begin{figure}[t]
\centering\includegraphics[width=8cm]{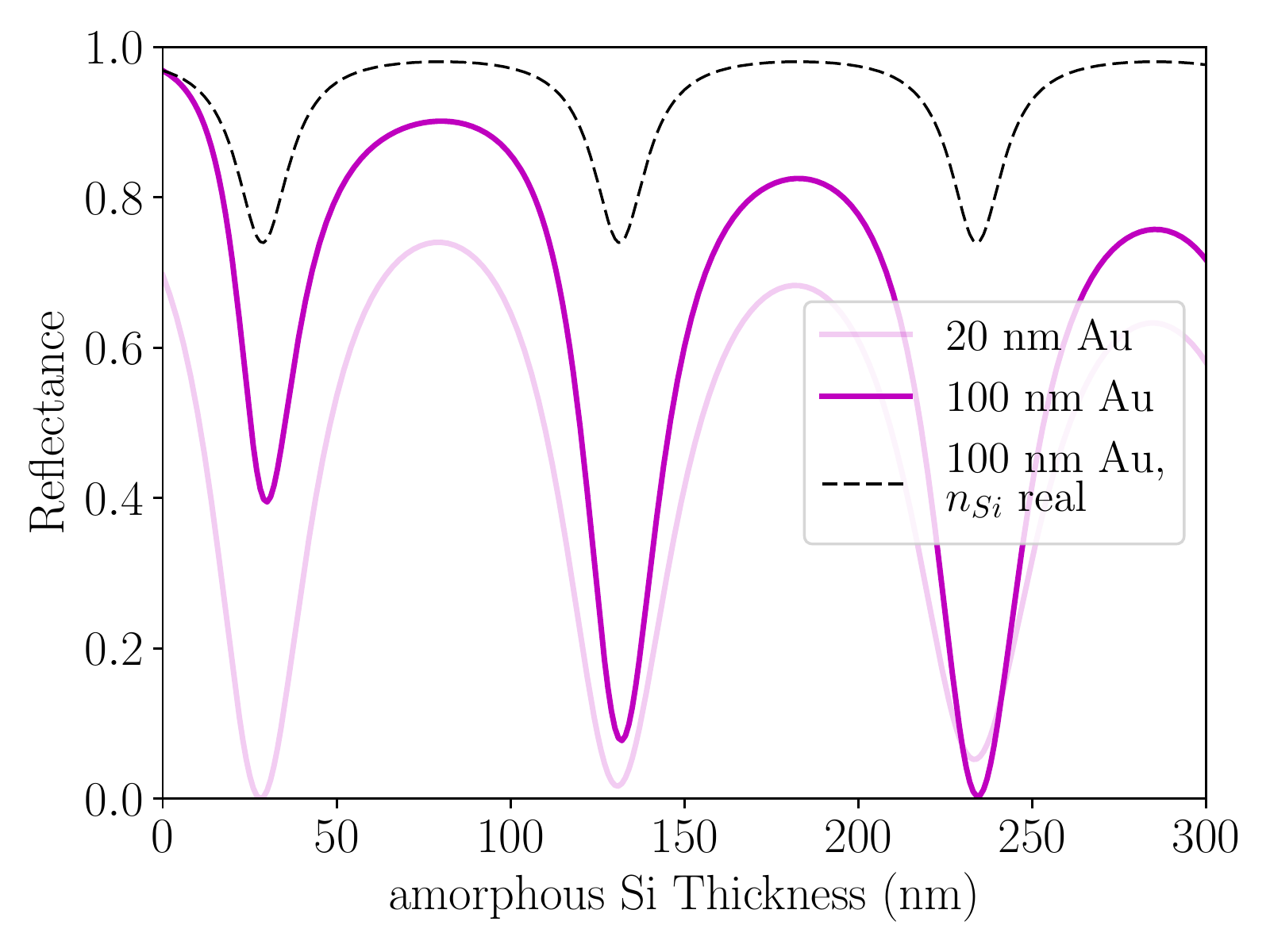}
\caption{ (Solid curves: Calculated reflectance at the wavelength of 800 nm for the Air/$\alpha$-Si/Au/Si(substrate)/Air multilayer, as a function of the $\alpha$-Si thickness, for two different Au thicknesses at normal incidence. Dashed curve: Calculated reflectance at a wavelength of 800 nm for an ideal dielectric on top of a 100 nm Au layer, characterized by $n=3.9$ and zero imaginary part of the refractive index.}
\label{fig2}
\end{figure}

We first used the transfer matrix method (TMM) \cite{bornwolf} to simulate the feasibility of this approach. We simulated the Air/$\alpha$-Si/Au/Si(substrate)/Air multilayer, where the outermost Air layers were semi-infinite, and the substrate was 500 $\mu$m thick. The radiation was assumed to be monochromatic with a wavelength of 800 nm, the typical center-wavelength of a Ti:sapphire laser, at a normal incidence to the multilayer stack. We used the refractive indices $n_{\rm Air}=1$, $n_{\alpha-\rm{Si}}\approx 3.90 + 0.11j$, $n_{\rm Au}\approx 0.15 + 4.91j$, and $n_{\rm Si}\approx 3.681 + 0.005j$ \cite{refrindex}.

In Fig.~\ref{fig2} we plot the reflectance of the stack as a function of the $\alpha$-Si thickness, for two gold layers with different thickness. For thin gold (20 nm), part of the radiation can be transmitted into the substrate, and $\approx30$ nm of amorphous silicon on top of it can efficiently suppress the reflectivity. For thick gold (100 nm), enough to prevent any transmission, a thicker amorphous silicon layer ($\approx230$ nm) is needed to achieve the same suppression.

In the same figure, we also plot the reflectivity (dashed lines) of a fictitious dielectric layer with zero imaginary part of the refractive index, and its magnitude equal to that of the the amorphous silicon, representing a conventional dielectric with negligible losses. It is evident that such a layer on top of a 100-nm thick gold layer cannot efficiently suppress the reflectivity, demonstrating that the cumulative losses after multiple reflections are necessary to realize the anti-reflection configuration.

We note that a single $\alpha$-Si anti-reflection coating remains efficient over a broad range of incidence angles. We used TMM to check that, when varying the incidence angle from 0 to 37.5 degrees, the optimal thickness for the $\alpha$-Si layer varies by less than 2\%. Furthermore, at the optimal thickness, the 800 nm reflectance remains below 0.05 (an acceptable value for the coating to properly work), at angles of incidence as high as 50 degrees. This can be understood as a consequence of the large refractive index of silicon, which causes the electromagnetic wave to strongly refract when entering the AR layer. As a result, the optical path in the silicon layer noticeably increases only at very large angles of incidence.

\section{Experimental and numerical verification}
\begin{figure}[t]
\centering\includegraphics[width=8cm]{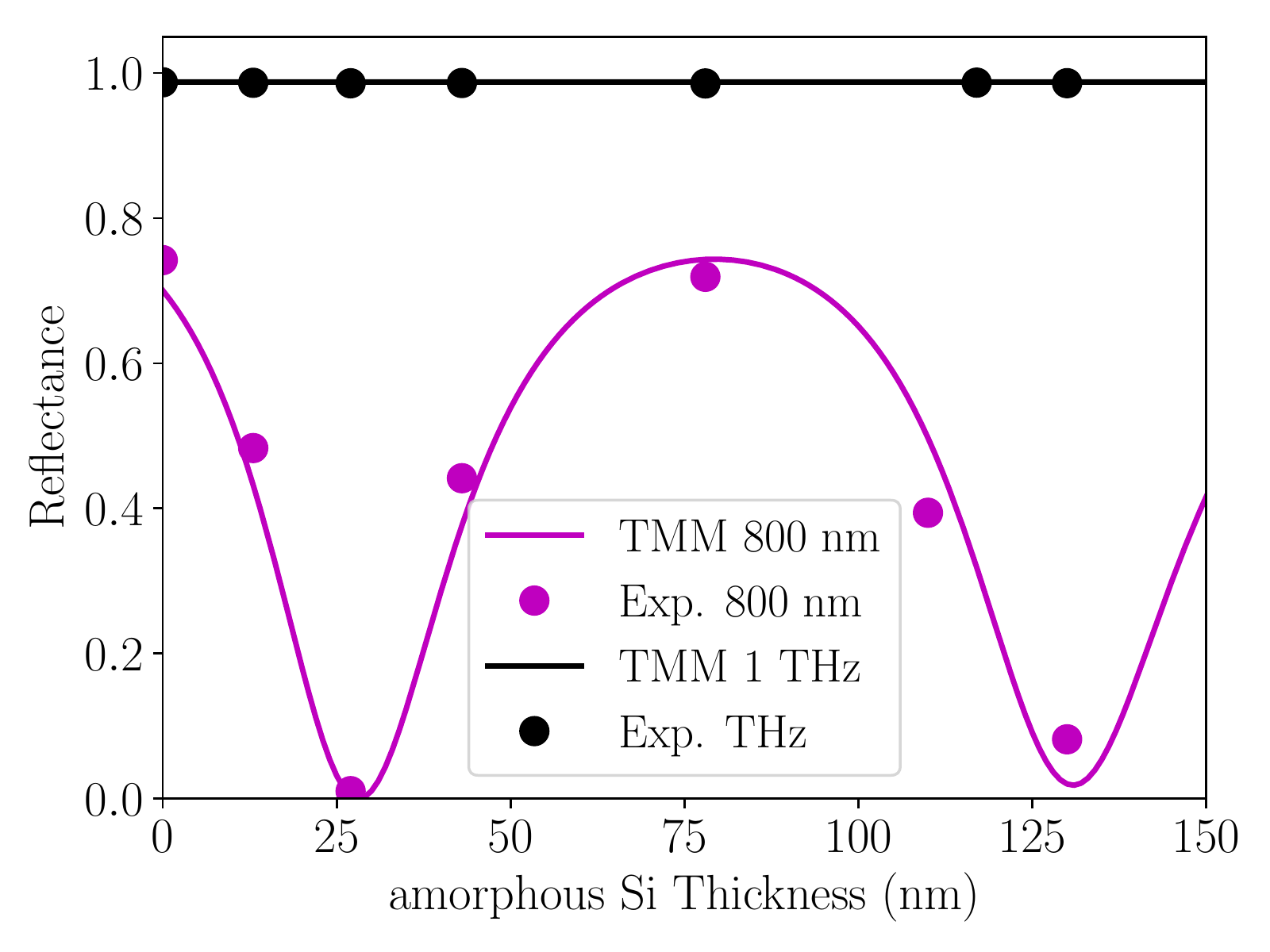}
\caption{Experimental (symbols) and calculated (line) reflectance for a $\alpha$-Si/Au/Si(substrate) sample based on a 20 nm-thick Au layer, at wavelengths of 800 nm (magenta) and 300 $\mu$m  (black), the latter corresponding to the radiation frequency of 1 THz.}
\label{fig3}
\end{figure}

In Fig.~\ref{fig3}, we plot the calculated and the measured reflectance for several Air/$\alpha$-Si($t$)/Au(20 nm)/Si(substrate)/Air multilayers, as a function of $t$, both for 800 nm and for the terahertz radiation impinging on the sample at 10 degrees incidence. The 800 nm radiation was produced by the Ti:sapphire-based regenerative amplifier (Coherent Legend) in  40 fs pulses, with the 30 nm FWHM bandwidth around the center 800 nm wavelength, as measured by a grating spectrometer. The reflectance at 800 nm was measured directly using a photodiode. The signal was scaled using the known reflectance value of a commercial gold mirror. The reflectance $R$ of the terahertz radiation, generated by optical rectification in a OH1 organic crystal \cite{oh1}, was determine from the measured transmittance $T$, using $R=1-A-T$, where the absorption $A$ was calculated based on the TMM. The transmittance was taken to be proportional to the square of the normalized amplitude of the maximum electro-optical sampling signal in a 100 $\mu$m thick, 110-cut GaP crystal.

The excellent agreement between the data and the calculations directly demonstrates the functionality of our design in suppressing the near-infrared reflectivity, with the appropriate thickness of $\alpha$-Si, 27 nm and 127 nm in the studied case of 800 nm radiation. We emphasize that our experiment demonstrates efficient suppression of the broadband 40 fs pulses of 800 nm radiation. This is not surprising, considering that the bandwidth to carrier ratio is less than 4\%. This result confirms the suitability of our design for the conventional ultrafast experiments. On the other hand, the terahertz reflectivity is unchanged by the $\alpha$-Si layer, suggesting that the terahertz near-field enhancement is also likely unaffected the silicon layer. However, since the measured reflectivity is a far-field property, and near-field properties are in general very sensitive to interface effects, one needs to perform a more detailed investigation of the possible effects of the anti-reflection coating on the terahertz radiation in the near-field regime.

\begin{figure}[t]
\centering\includegraphics[width=8cm]{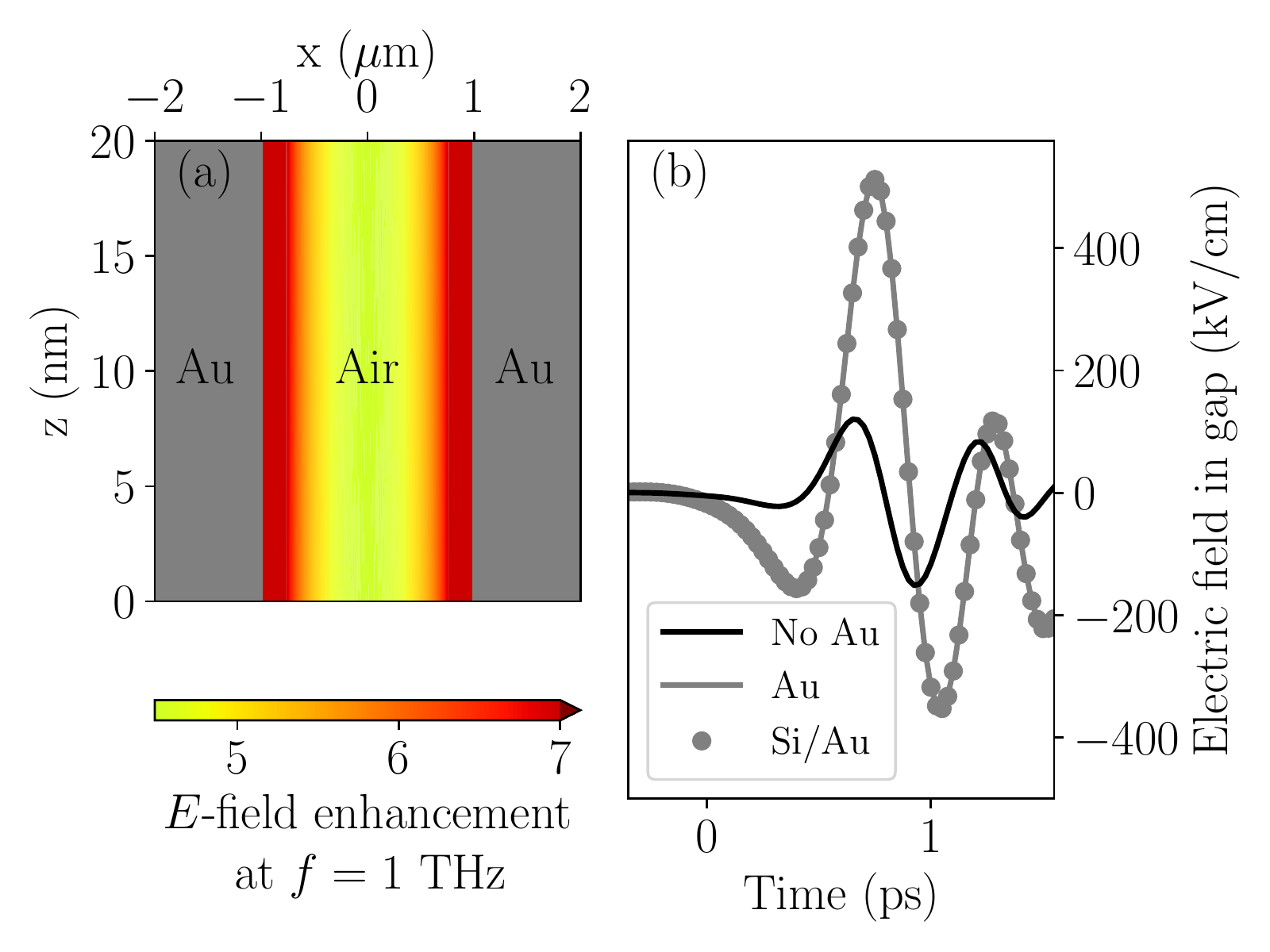}
\caption{(a) Frequency-domain, finite element analysis of the enhancement map for a monochromatic electromagnetic field with frequency $f=1$ THz incident on two gold plates separated by a gap. (b) Time-domain finite element simulations of the $x$-component of the electric field for a single-cycle (broadband) terahertz field at the center of the gap, without  Au plates (solid black curve), with Au plates (solid gray curve) and with $\alpha$-Si/Au plates (filled gray dots). In all the calculations, the electric field of the incident radiation is polarized along the $x-$axis, and the propagation direction is along the $z-$axis, normal to the sample plane.}
\label{fig4}
\end{figure}

To analyze the near-field effects of the coating, we have performed finite-element numerical calculations using COMSOL Multiphysics\textregistered software~\cite{comsol}. In Fig.~\ref{fig4}(a), we plot the electric field enhancement at a frequency of 1 THz, for a set of two infinitely long, 65 $\mu$m wide, 20 nm thick gold plates separated by a gap of 2 $\mu$m. The terahertz electric field is applied along the $x-$axis in this Figure. The field enhancement is computed by dividing the electric field value in the gap region by the electric field value at the air/silicon interface, in the absence of the gold plates. In Fig.~\ref{fig4}(b), we plot time-dependence of the $x$-component of the terahertz electric field in the middle of the gap ($x=0$), and at $z=10$ nm above the silicon substrate. The simulation used the experimental time profile of the impinging terahertz field,  measured by electro-optical sampling, with the peak value of $\approx300$ kV/cm.

For the bare silicon-air interface, the terahertz field is reduced to approximately half of its free-space magnitude, consistent with the relative amplitude of the transmitted wave at an air/silicon interface computed as $t=2/(n+1)$, with $n\approx3.4$ \cite{footnote3}. The presence of the gold plates introduces a slight temporal shift, and enhances the amplitude of the terahertz field to more than 500 kV/cm, consistent with the $\approx4$ times enhancement observed in the frequency domain simulations of Fig.~\ref{fig4}(a) \cite{footnote2}. Most importantly, the addition of the $\alpha$-Si layer on top of the gold plates does not noticeably affect the field, confirming our intuitive conclusions based on the negligible effect of the $\alpha$-Si layer on the terahertz reflectivity.

To test the functionality of the AR coating in a practical configuration, we measured the polar magneto-optical Kerr effect (MOKE) \cite{qiu00} from the 3 nm-thick CoNi film patterned into a 1 $\mu$m-wide, 100 $\mu$m-long wire. The CoNi stack is formed by a Ta(2)|Cu(2)|[Co(0.2)|Ni(1)]$_{3}$|Ni(0.5)|Ta(3) multilayer (thicknesses in nm) with perpendicular magnetic anisotropy. The wire is located in the 2 $\mu$m-wide gap between two 100 $\mu$m long and 65 $\mu$m wide gold plates, coated with 27 nm of $\alpha$-Si. By analyzing the magneto-optic response, we can unambiguously identify the signal coming from the embedded CoNi wire, with no contribution from the non-magnetic electrodes. As the MOKE signal typically results in a tiny intensity variation on top of a large background, this specific system implementation also demonstrates the general suitability of our design for the detection of small effects other than magneto-optical ones.

\begin{figure}[t]
\centering\includegraphics[width=8cm]{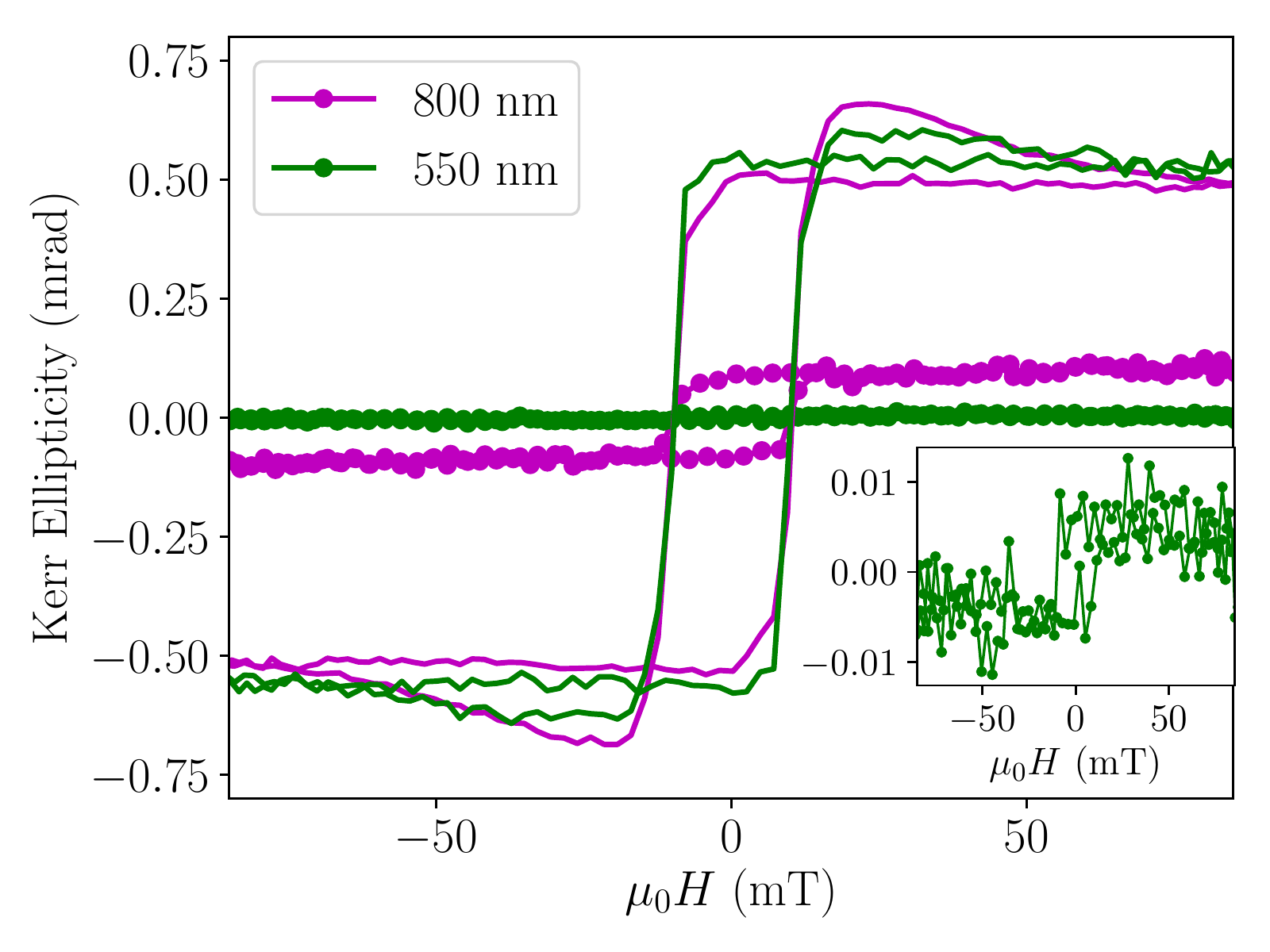}
\caption{Polar Kerr ellipticity as a function of the magnetic field applied orthogonal to the sample plane, for the 800 nm and 550 nm wavelengths of the probing light. Symbols: average of 25 hysteresis loops for a  1 $\mu$m-wide, 100 $\mu$m-long CoNi wire. Solid curves:  average of 4 hysteresis loops hysteresis loops for a 100 $\mu$m $\times$ 100 $\mu$m CoNi square, at  the same wavelengths. Inset: Zoom in on the 550 nm wavelength hysteresis loop for the CoNi wire.}
\label{fig5}
\end{figure}

The polar MOKE loops from the embedded wire are plotted in the main panel of Fig.~\ref{fig5} for two different wavelengths of the probing radiation, 550 nm and 800 nm. At 800 nm, the AR coating optimized for this wavelength is expected to completely suppress the reflectivity of the gold electrodes, while at 550 nm, a substantial reflection from the metallic pads is expected. Using radiation of different wavelengths is geometrically equivalent to studying samples with different AR coating thickness, with the advantage that the very same sample can be used and the wavelength can be tuned very accurately.

The plotted MOKE signals reflect the change in the polarization ellipticity of the probing light, determined with a suitable polarization analyzing system. In these measurements, we utilized a polarization modulation technique \cite{vavassori00} to provide the sensitivity necessary for testing the effectiveness of the proposed AR coating. In this setup, the polarization of the incident light was modulated at the frequency $\omega$, while both the total measured intensity $I_0$ and its variation $I_\omega$ at the modulation frequency 
were simultaneously recorded. The polarization ellipticity was determined from $I_\omega$ normalized by the total intensity $I_0$ reflected from the sample. While the variations of $I_\omega$ were affected only by the magnetic structure, the magnitude of $I_0$ is determined by the whole probed area, including the gold plates.

The data plotted with symbols in Fig.~\ref{fig5} clearly show that the AR coating significantly enhances the signal-to-background ratio, resulting in a more than a ten-fold increase of the realtive amplitude of the loop at the design wavelength. We have checked that the increase in the relative signal is not caused by the difference between the magneto-optical constants of CoNi between 800 and 550 nm wavelengths, by measuring the MOKE signal from a 100 $\mu$m $\times$ 100 $\mu$m CoNi square of 3 nm thickness, with no gold electrodes surrounding the structure. The resulting hysteresis loops are shown with solid curves in the the same Fig.~\ref{fig5}. 

To quantitatively analyze our observations, we note that if the areas and the reflectivity of different reflecting regions are known, one can predict the difference in the total measured Kerr ellipticity $\epsilon_K$ between the CoNi square larger than the probing spot, and the wire-shaped sample, according to \cite{qureshi05}
\begin{equation}
\dfrac{\epsilon_{K, \rm wire}}{\epsilon_{K, \rm square}} = \dfrac{A_{\rm wire}}{A_{\rm wire} + A_{\rm Si}\dfrac{R_{\rm Si}}{R_{\rm CoNi}} + A_{\rm CoatedAu}\dfrac{R_{\rm CoatedAu}}{R_{\rm CoNi}}},
\label{eq_model}
\end{equation}
where $A_m$ is the total area occupied by a certain material $m$ illuminated by the laser beam, and $R_m$ the corresponding reflectivity that can be measured experimentally or calculated using Fresnel equations.

Table~\ref{table1} summarizes the relationship between the ellipticity of the wire and the ellipticity of the square, assuming that the probing light is focused in to a uniform circular spot with diameter $\phi=75$ $\mu$m, and the various probed areas are $A_{\rm wire}=A_{\rm Si}\approx\phi h$ $(h=1~\mu\rm m)$, $A_{\rm CoatedAu}\approx\pi(\phi/2)^2-2\phi h$. The area occupied by the wire is therefore about 2\% of the total area. Indeed, for the wavelength of 550 nm, at which the Au reflectivity is not suppressed by the AR coating, the ellipticity signal for the wire is about 1.6\% of that for the large square, both theoretically and experimentally. In contrast, for the 800 nm wavelength, we expect and observe an increase of this ratio by an order of magnitude. The deviation between the theoretically expected value (24\%) and the experimental value (19\%) can be explained by a combination of a few-nm uncertainty in the deposited material thickness, deviations of the optical properties of different layers from their nominal values, and by the effects of the nanowire and gold electrode edges, whose scattering properties were not taken into account in the calculations reported in Table~\ref{table1}.

\begin{table}[t]
\caption{Summary of the ellipticity ratio between a CoNi wire and a CoNi square calculated according to Eq.~(\ref{eq_model}).}
\begin{center}
    \begin{tabular}{| l | c | c | c | c |}
    \hline
     &  $R_{\rm CoNi}$ & $R_{\rm Si}$ & $R_{\rm CoatedAu}$ & $\epsilon_{K, \rm wire}/\epsilon_{K, \rm square}$\\ \hline \hline
    550 nm (theor.) & 0.49 & 0.40 & 0.50 & $\mathbf{0.017}$\\ \hline
    550 nm (exp.) & 0.46 & 0.40 & 0.50 & $\mathbf{0.016}$\\ \hline\hline
    800 nm (theor.) & 0.47 & 0.36 & 0.020 & $\mathbf{0.24}$\\ \hline
    800 nm (exp.) & 0.49 & 0.34 & 0.031 & $\mathbf{0.19}$\\ \hline
    \end{tabular}
\end{center}
\label{table1}
\end{table}

\section{Conclusion}
In summary, we have designed and experimentally demonstrated an anti-reflection coating for highly reflective metals, which are typically utilized in the fabrication of terahertz meta-materials. The anti-reflection coating can efficiently suppress the reflection of light in the visible and infrared ranges, typically used in the studies of ultrafast phenomena by  pump-probe techniques. At the same time, the coating does not perturb the propagation of terahertz radiation, and does not affect the near-field enhancement in the meta-materials. Our results are expected to open a path for time-resolved experiments aimed at probing the ultrafast dynamics driven in nano-scale structures by strong terahertz fields, by using table-top femtosecond near-infrared laser sources.


\section*{Funding}
Swedish Research Council (Grant E0635001); Marie Sk\l{}odowska Curie Actions, Cofund (Project INCA 600398s); European Research Council (ERC) (Starting Grant 715452 ``MAGNETIC-SPEED-LIMIT''); H2020 European Union Programme, FETOPEN-2016-2017 (Project 737093 ``FEMTOTERABYTE'');  Spanish Ministry of Economy, Industry and Competitiveness under the Maria de Maeztu Units of Excellence Programme - MDM-2016-0618; US National Science Foundation (NSF) (Grant ECCS-1503878, Grant DMR-1504449); U.S. Department of Energy, Office of Science, Office of Basic Energy Sciences, Award No. 2015-SLAC-100238-Funding.




\begin{thebibliography}{10}

\bibitem{hoffmann11}
M. C. Hoffmann and J. A. F\"ul\"op, ``Intense ultrashort terahertz pulses:
  generation and applications ,'' J. Phys. D: Appl. Phys. {\bfseries 44}(8),
  083001 (2011).

\bibitem{trigo08}
M. Trigo, Y. M. Sheu, D. A. Arms, J. Chen, S. Ghimire, R. S. Goldman, E.
  Landahl, R. Merlin, E. Peterson, M. Reason, and D. A. Reis, ``Probing
  Unfolded Acoustic Phonons with X Rays,'' \prl {\bfseries 101}(2), 025505
  (2008).

\bibitem{liu12}
M. Liu, H. Y. Hwang, H. Tao, A. C. Strikwerda, K. Fan, G. R. Keiser, A. J.
  Sternbach, K. G. West, S. Kittiwatanakul, J. Lu, S. A. Wolf, F. G. Omenetto,
  X. Zhang, K. A. Nelson, and R. D. Averitt, ``Terahertz-field-induced
  insulator-to-metal transition in vanadium dioxide metamaterial,'' \nat
  {\bfseries 487}(7407), 345--348 (2012).

\bibitem{daranciang12}
D. Daranciang, M. J. Highland, H. Wen, S. M. Young, N. C. Brandt, H. Y. Hwang,
  M. Vattilana, M. Nicoul, F. Quirin, J. Goodfellow, T. Qi, I. Grinberg, D. M.
  Fritz, M. Cammarata, D. Zhu, H. T. Lemke, D. A. Walko, E. M. Dufresne, Y. Li,
  J. Larsson, D. A. Reis, K. Sokolowski-Tinten, K. A. Nelson, A. M. Rappe, P.
  H. Fuoss, G. B. Stephenson, and A. M. Lindenberg, ``Ultrafast Photovoltaic
  Response in Ferroelectric Nanolayers,'' \prl {\bfseries 108}(8), 087601
  (2012).

\bibitem{mankowsky14}
R. Mankowsky, A. Subedi, M. F\"orst, S. O. Mariager, M. Chollet, H. T. Lemke,
  J. S. Robinson, J. M. Glownia, M. P. Minitti, A. Frano, M. Fechner, N. A.
  Spaldin, T. Loew, B. Keimer, A. Georges, and A. Cavalleri, ``Nonlinear
  lattice dynamics as a basis for enhanced superconductivity in
  YBa$_2$Cu$_3$O$_{6.5}$,'' \nat {\bfseries 516}(7529), 71--73 (2014).

\bibitem{staub14}
U. Staub, R. A. de Souza, P. Beaud, E. M\"ohr-Vorobeva, G. Ingold, A. Caviezel,
  V. Scagnoli, B. Delley, W. F. Schlotter, J. J. Turner, O. Krupin, W.-S. Lee,
  Y.-D. Chuang, L. Patthey, R. G. Moore, D. Lu, M. Yi, P. S. Kirchmann, M.
  Trigo, P. Denes, D. Doering, Z. Hussain, Z. X. Shen, D. Prabhakaran, A. T.
  Boothroyd, and S. L. Johnson, ``Persistence of magnetic order in a highly
  excited Cu$^{2+}$ state in CuO,'' \prb {\bfseries 89}(22), 220401(R) (2014).

\bibitem{beaud14}
P. Beaud, A. Caviezel, S. O. Mariager, L. Rettig, G. Ingold, C. Dornes, S.-W.
  Huang, J. A. Johnson, M. Radovic, T. Huber, T. Kubacka, A. Ferrer, H. T.
  Lemke, M. Chollet, D. Zhu, J. M. Glownia, M. Sikorski, A. Robert, H. Wadati,
  M. Nakamura, M. Kawasaki, Y. Tokura, S. L. Johnson, and U. Staub, ``A
  time-dependent order parameter for ultrafast photoinduced phase
  transitions,'' Nat. Mater. {\bfseries 13}(10), 923--927 (2014).

\bibitem{kubacka14}
T. Kubacka, J. A. Johnson, M. C. Hoffmann, C. Vicario, S. de Jong, P. Beaud, S.
  Gr\"ubel, S.-W. Huang, L. Huber, L. Patthey, Y.-D. Chuang, J. J. Turner, G.
  L. Dakovski, W.-S. Lee, M. P. Minitti, W. Schlotter, R. G. Moore, C. P.
  Hauri, S. M. Koohpayeh, V. Scagnoli, G. Ingold, S. L. Johnson, and U. Staub,
  ``Large-Amplitude Spin Dynamics Driven by a THz Pulse in Resonance with an
  Electromagnon,'' Science {\bfseries 343}(6177), 1333--1336 (2014).

\bibitem{dakovski15}
G. L. Dakovski, W.-S. Lee, D. G. Hawthorn, N. Garner, D. Bonn, W. Hardy, R.
  Liang, M. C. Hoffmann, and J. J. Turner, ``Enhanced coherent oscillations in
  the superconducting state of underdoped YBa$_2$Cu$_3$O$_{6+x}$ induced via
  ultrafast terahertz excitation,'' \prb {\bfseries 91}(22), 220506(R) (2015).

\bibitem{bonetti16}
S. Bonetti, M. C. Hoffmann, M.-J. Sher, Z. Chen, S.-H. Yang, M. G. Samant, S.
  S. P. Parkin, and H. A. D\"urr, ``THz-Driven Ultrafast Spin-Lattice
  Scattering in Amorphous Metallic Ferromagnets,'' \prl {\bfseries 117}(8),
  087205 (2016).

\bibitem{dean16}
M. P. M. Dean, Y. Cao, X. Liu, S. Wall, D. Zhu, R. Mankowsky, V. Thampy, X. M.
  Chen, J. G. Vale, D. Casa, J. Kim, A. H. Said, P. Juhas, R. Alonso-Mori, J.
  M. Glownia, A. Robert, J. Robinson, M. Sikorski, S. Song, M. Kozina, H.
  Lemke, L. Patthey, S. Owada, T. Katayama, M. Yabashi, Y. Tanaka, T. Togashi,
  J. Liu, C. Rayan Serrao, B. J. Kim, L. Huber, C.-L. Chang, D. F. McMorrow, M.
  F\"orst, and J. P. Hill, ``Ultrafast energy- and momentum-resolved dynamics
  of magnetic correlations in the photo-doped Mott insulator Sr$_2$IrO$_4$,''
  Nat. Mater. {\bfseries 15}(6), 601--605 (2016).

\bibitem{henighan16}
T. Henighan, M. Trigo, M. Chollet, J. N. Clark, S. Fahy, J. M. Glownia, M. P.
  Jiang, M. Kozina, H. Liu, S. Song, D. Zhu, and D. A. Reis, ``Control of
  two-phonon correlations and the mechanism of high-wavevector phonon
  generation by ultrafast light pulses,'' \prb {\bfseries 94}(2), 020302(R)
  (2016).

\bibitem{grubel16}
S. Gr\"ubel, J. A. Johnson, P. Beaud, C. Dornes, A. Ferrer, V. Haborets, L.
  Huber, T. Huber, A. Kohutych, T. Kubacka, M. Kubli, S. O. Mariager, J.
  Rittmann, J. I. Saari, Y. Vysochanskii, G. Ingold, and S. L. Johnson,
  ``Ultrafast x-ray diffraction of a ferroelectric soft mode driven by
  broadband terahertz pulses,'' arXiv:1602.05435v1 (2016).

\bibitem{chen16}
F. Chen, Y. Zhu, S. Liu, Y. Qi, H. Y. Hwang, N. C. Brandt, J. Lu, F. Quirin, H.
  Enquist, P. Zalden, T. Hu, J. Goodfellow, M.-J. Sher, M. C. Hoffmann, D. Zhu,
  H. Lemke, J. Glownia, M. Chollet, A. R. Damodaran, J. Park, Z. Cai, I. W.
  Jung, M. J. Highland, D. A. Walko, J. W. Freeland, P. G. Evans, A. Vailionis,
  J. Larsson, K. A. Nelson, A. M. Rappe, K. Sokolowski-Tinten, L. W. Martin, H.
  Wen, and A. M. Lindenberg, ``Ultrafast terahertz-field-driven ionic response
  in ferroelectric BaTiO$_3$,'' \prb {\bfseries 94}(18), 180104(R) (2016).

\bibitem{chen06}
H.-T. Chen, W. J. Padilla, J. M. O. Zide, A. C. Gossard, A. J. Taylor, and R.
  D. Averitt, ``Active terahertz metamaterial devices,'' \nat {\bfseries
  444}(7119), 597--600 (2006).

\bibitem{razzari11}
L. Razzari, A. Toma, M. Shalaby, M. Clerici, R. Proietti Zaccaria, C. Liberale,
  S. Marras, I. A. I. Al-Naib, G. Das, F. De Angelis, M. Peccianti, A. Falqui,
  T. Ozaki, R. Morandotti, and E. Di Fabrizio, ``Extremely large extinction
  efficiency and field enhancement in terahertz resonant dipole nanoantennas,''
  \opex {\bfseries 19}(27), 26088--26094 (2011).

\bibitem{werley12}
C. A. Werley, K. Fan, A. C. Strikwerda, S. M. Teo, X. Zhang, R. D. Averitt, and
  K. A. Nelson, ``Time-resolved imaging of near-fields in THz antennas and
  direct quantitative measurement of field enhancements,'' \opex {\bfseries
  20}(8), 8551--8567 (2012).

\bibitem{zhang15}
J. Zhang, X. Zhao, K. Fan, X. Wang, G.-F. Zhang, K. Geng, X. Zhang, and R. D.
  Averitt, ``Terahertz radiation-induced sub-cycle field electron emission
  across a split-gap dipole antenna,'' \apl {\bfseries 107}(23), 231101 (2015).

\bibitem{savoini16}
M. Savoini, S. Gr\"ubel, S. Bagiante, H. Sigg, T. Feurer, P. Beaud, and S. L.
  Johnson, ``THz near-field enhancement by means of isolated dipolar antennas:
  the effect of finite sample size,'' \opex {\bfseries 24}(5), 4552--4562
  (2016).

\bibitem{kozina17}
M. Kozina, M. Pancaldi, C. Bernhard, T. van Driel, J.M. Glownia, P. Marsik, M.
  Radovic, C. A. F. Vaz, D. Zhu, S. Bonetti, U. Staub, and M.C. Hoffmann,
  ``Local Terahertz Field Enhancement for Time-Resolved X-ray Diffraction,''
  \apl {\bfseries 110}(8), 081106 (2017).

\bibitem{balasubramanian88}
K. Balasubramanian, A. S. Marathay, and H. A. Macleod, ``Modeling
  magneto-optical thin film media for optical data storage,'' Thin Solid Films
  {\bfseries 164}, 391--403 (1988).

\bibitem{atkinson91}
R. Atkinson, I. W. Salter, and J. Xu, ``Quadrilayer magneto-optic enhancement
  with zero Kerr ellipticity,'' J. Magn. Magn. Mater. {\bfseries 102}(3),
  357--364 (1991).

\bibitem{qureshi04}
N. Qureshi, H. Schmidt, and A. R. Hawkins, ``Cavity enhancement of the
  magneto-optic Kerr effect for optical studies of magnetic nanostructures,''
  \apl {\bfseries 85}(3), 431 (2004).

\bibitem{qureshi05}
N. Qureshi, S. Wang, M. A. Lowther, A. R. Hawkins, S. Kwon, A. Liddle, J.
  Bokor, and H. Schmidt, ``Cavity-Enhanced Magnetooptical Observation of
  Magnetization Reversal in Individual Single-Domain Nanomagnets,'' Nano Lett.
  {\bfseries 5}(7), 1413--1417 (2005).

\bibitem{barman06}
A. Barman, S. Wang, J. D. Maas, A. R. Hawkins, S. Kwon, A. Liddle, J. Bokor,
  and H. Schmidt, ``Magneto-Optical Observation of Picosecond Dynamics of
  Single Nanomagnets,'' Nano Lett. {\bfseries 6}(12), 2939--2944 (2006).

\bibitem{wang07}
S. Wang, A. Barman, H. Schmidt, J. D. Maas, A. R. Hawkins, S. Kwon, B.
  Harteneck, S. Cabrini, and J. Bokor, ``Optimization of nano-magneto-optic
  sensitivity using dual dielectric layer enhancement,'' \apl {\bfseries
  90}(25), 252504 (2007).

\bibitem{biagioni12}
P. Biagioni, J.-S. Huang, and B. Hecht, ``Nanoantennas for visible and infrared
  radiation,'' Rep. Prog. Phys. {\bfseries 75}(2), 024402 (2012).

\bibitem{hass56}
G. Hass, H. H. Schroeder, and A. F. Turner, ``Mirror Coatings for Low Visible
  and High Infrared Reflectance,'' \josa {\bfseries 46}(1), 31--35 (1956).

\bibitem{yoshida79}
S. Yoshida, ``Antireflection coatings on metals for selective solar
  absorbers,'' Thin Solid Films {\bfseries 56}(3), 321--329 (1979).

\bibitem{park64}
K. C. Park, ``The Extreme Values of Reflectivity and the Conditions for Zero
  Reflection from Thin Dielectric Films on Metal,'' \ao {\bfseries 3}(7),
  877--881 (1964).

\bibitem{adachi00}
S. Adachi and H. Mori, ``Optical properties of fully amorphous silicon,'' \prb
  {\bfseries 62}(15), 10158 (2000).

\bibitem{bornwolf}
M. Born and E. Wolf, \textit{Principles of Optics} (Cambridge University,
  1999).

\bibitem{refrindex}
M. N. Polyanskiy, ``Refractive index database,''
  \url{https://refractiveindex.info}.

\bibitem{oh1}
F. D. J. Brunner, O-P. Kwon, S.-J. Kwon, M. Jazbin\v{s}ek, A. Schneider, and P.
  G\"unter, ``A hydrogen-bonded organic nonlinear optical crystal for
  high-efficiency terahertz generation and detection,'' \opex {\bfseries
  16}(21), 16496--16508 (2008).

\bibitem{comsol}
COMSOL Multiphysics\textregistered\ v. 5.3. \url{https://www.comsol.com}.
  COMSOL, AB, Stockholm, Sweden.

\bibitem{footnote3}
Since we are impinging on the air/silicon interface at normal incidence with a
  wavelength of $\approx300$ $\mu$m, the amplitude of the electric field at
  $z=10$ nm above the silicon substrate is almost equal to the amplitude of the
  transmitted electric field, because of the $\hat{n} \times \left(
  \vec{E}_{Si} - \vec{E}_{Air} \right) = 0$ boundary condition.

\bibitem{footnote2}
The discrepancy between the two values has to do with the fact that the
  single-cycle field is broadband, and different frequencies are amplified
  differently by a fixed-geometry design.

\bibitem{qiu00}
Z. Q. Qiu and S. D. Bader, ``Surface magneto-optic Kerr effect,'' Rev. Sci.
  Instrum. {\bfseries 71}(3), 1243 (2000).

\bibitem{vavassori00}
P. Vavassori, ``Polarization modulation technique for magneto-optical
  quantitative vector magnetometry,'' \apl {\bfseries 77}(11), 1605 (2000).

\end{thebibliography}
\end{document}